\documentclass[11pt]{article}
\usepackage[margin=0.82in]{geometry}
\usepackage[T1]{fontenc}
\usepackage[utf8]{inputenc}
\usepackage{newtxtext,newtxmath}
\usepackage{amsmath,bm}
\usepackage{graphicx}
\usepackage{booktabs}
\usepackage{microtype}
\usepackage{xcolor}
\usepackage{caption}
\usepackage{float}
\usepackage{hyperref}
\usepackage[backend=biber,style=numeric-comp,sorting=none,maxnames=6,minnames=3,giveninits=true]{biblatex}
\hypersetup{colorlinks=true,linkcolor=black,citecolor=black,urlcolor=blue,
  pdftitle={A regenerating free-energy register for protein-templated period-2 DNA synthesis by Drt3b},
  pdfsubject={Biological Physics; Drt3b generating-register model},
  pdfkeywords={Drt3b, protein-templated DNA synthesis, semi-Markov, no-refit validation},
  pdfauthor={Wei-Wei Zhang}}
\newcommand{\dd}{\mathrm d}

\begin{document}
\begin{center}
{\Large\bfseries A regenerating free-energy register for\\
protein-templated period-2 DNA synthesis by Drt3b\par}
\vspace{0.22em}
{\large A locked electronic-to-event framework for sequence, timing, and perturbation prediction\par}
\vspace{0.65em}
{\bfseries Wei-Wei Zhang, MD, PhD}\par
Adventin Inc., San Diego, California, USA\par
Correspondence: \href{mailto:wzhang@adventin.com}{wzhang@adventin.com}\par
\end{center}

\begin{abstract}
Deng et al. discovered that Drt3b synthesizes protein-primed poly(AC) DNA without a nucleic-acid template. The structures reveal a finite protein architecture and product-associated contacts, but not the renewable dynamical rule that converts a bounded pocket into a long period-2 sequence. We propose that Drt3b is a regenerating protein--substrate free-energy register. The framework specifies a sampled quantum mechanics/molecular mechanics (QM/MM) potential of mean force and a preregistered, within-domain first-order electronic-descriptor map for edge-specific activation barriers; physical curvature is admitted only through a training-frozen Taylor-remainder bound that widens uncertainty but never corrects a held-out mean. Conditional hazards derived from those barriers are averaged over unresolved conformational, hydration, protonation, and metal-coordination trajectories to yield a semi-Markov event kernel. Pointwise ratios of competing hazards are therefore fixed by the same barrier differences and preregistered prefactor rules that govern exit timing, so nucleotide choice and dwell statistics are co-generated rather than separately fitted. A regime gate separates stationary or slowly driven experiments, analyzed by a joint sequence--time cycle determinant, from genuinely nonstationary protocols, analyzed by the full age-structured tilted propagator. The decisive test is one locked parameterization: without condition-specific refitting, it must jointly predict nucleotide choice, waiting-time laws, product-length tails, and the rank, minors, null spaces, and rescue structure of perturbation responses. An independent 2026 DRT3 study supplies a stringent cross-construct transport test after sequence and structural alignment are frozen. The proposal requires neither coherent quantum computation nor coupling to an external field.
\end{abstract}

\section*{The unresolved generative problem}
Deng et al. reported a $D_3$-symmetric $6{:}6{:}6$ Drt3a--Drt3b--ncRNA assembly. Drt3a copies an RNA AC repeat to make poly(GT), whereas Drt3b produces the complementary poly(AC) strand without a nucleic-acid template and initiates the product covalently from Tyr650 \cite{Deng2026,PDB9Z6Y,PDB9Z6Z}. The deposited elongating and resting structures are reported at 2.6~\AA\ resolution and contain product-associated Glu26--dA and Arg253--dC contacts. These observations support a compact protein register, but endpoint reconstructions do not resolve transient rotamer, hydration, protonation, metal-coordination, product-end, or pre-chemistry states.

A second 2026 study independently reports a hexameric DRT3b that performs protein-primed, amino-acid-gated poly(dCdA) synthesis, with residues E22, R241, and Y666 in a distinct construct and numbering scheme \cite{Wang2026DRT3}. The two studies strengthen the biological premise while creating a demanding test: residue labels must not be merged by analogy. Sequence and structural alignment must be frozen before any cross-construct response comparison.

A functional template must discriminate among monomers, align reactive geometry, support chemical commitment, and renew the boundary condition for the next event. Rev1 side-chain templating, bacteriophage $\phi29$ protein priming, CCA addition, and terminal transferase activity show that these functions can be distributed across protein, product, and substrate \cite{Nair2005,Kamtekar2006,Mendez1992,Li2002CCA,Shi1998,Delarue2002}. Drt3b is unusual because priming, protein-associated base selection, and sustained period-2 writing coexist. A mechanism that fits the final sequence but cannot predict event timing and perturbation propagation is not yet generative.

\newpage
\begin{figure}[H]
\centering
\includegraphics[width=0.98\textwidth]{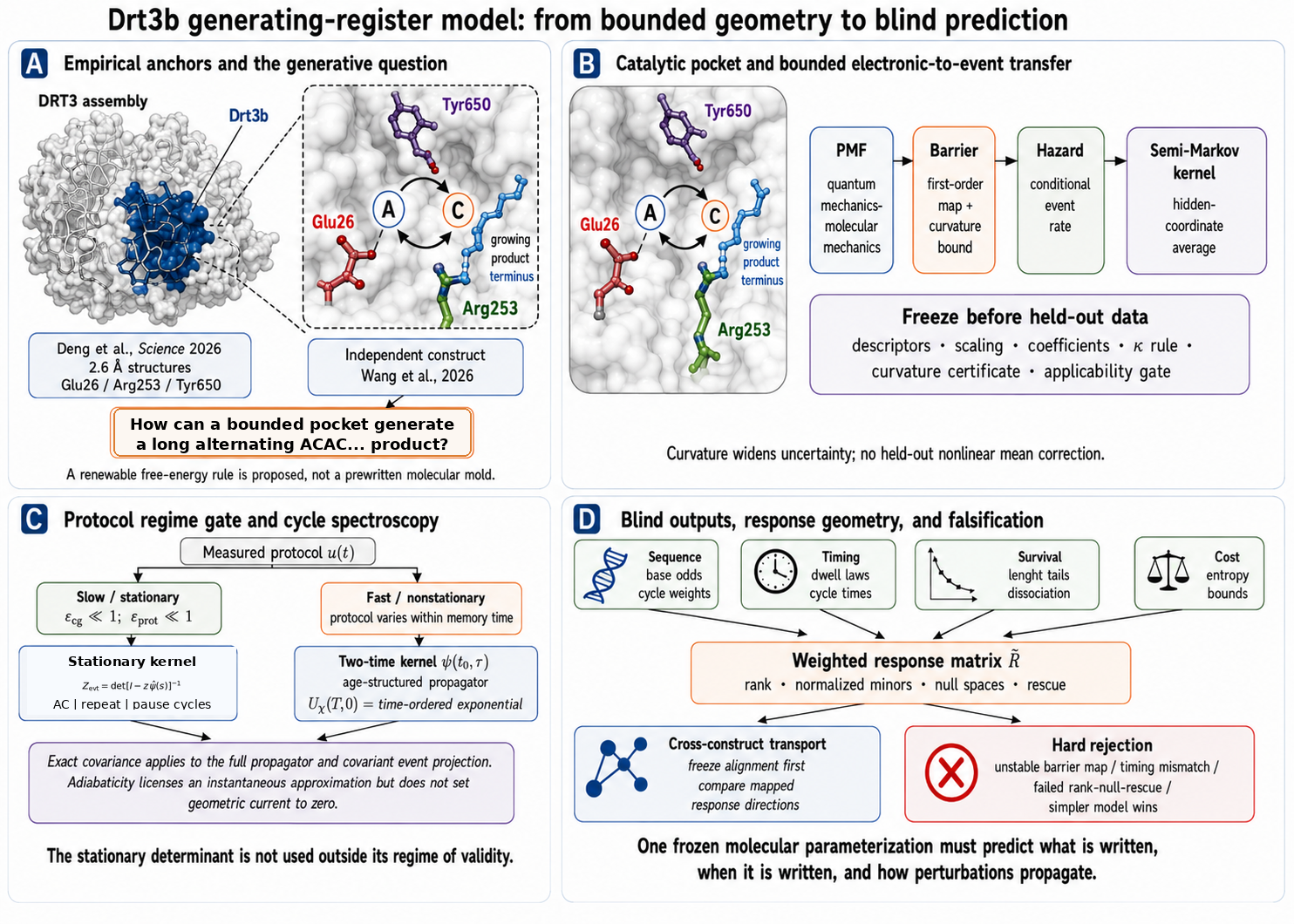}
\caption{Drt3b generating-register model: from bounded geometry to blind prediction. (A) Published DRT3 structures and an independent construct anchor the finite architecture and define the generative question. (B) A sampled potential of mean force is transferred through a frozen first-order barrier map, conditional hazard, and semi-Markov event kernel. (C) A protocol gate separates stationary cycle spectroscopy from a nonstationary two-time propagator. (D) One frozen molecular parameterization must predict sequence, timing, survival, and perturbation-response geometry on held-out data. Structural panels are original mechanistic schematics anchored to 9Z6Y/9Z6Z and reported residue contacts, not coordinate-exact renderings.}
\label{fig:overview}
\end{figure}

\section*{A locked electronic-to-event map}
For register state $r$, candidate nucleotide $b$, controlled perturbation $u$, reaction coordinate $\xi$, and remaining coordinates $Y$, define the sampled potential of mean force
\begin{equation}
G_{r,b}(\xi;u)=-k_BT\ln\!\int \dd Y\,\exp[-\beta H_{\rm QM/MM}(\xi,Y\mid r,b,u)]+C.
\label{eq:pmf-main}
\end{equation}
The activation free energy includes electronic discrimination, metal and proton coordinates, solvation, protein reorganization, and entropy. Trajectory decomposition may attribute barrier changes to electrostatics, exchange, polarization, dispersion, charge transfer, active-site fields, hydrogen bonding, metal coordination, and hydration, but it does not replace the sampled free energy \cite{Warshel1976,TorrieValleau1977,Pan2023,Xiong2024}.

The primary transfer model is deliberately local and constrained. Define the scaled descriptor displacement
\begin{equation}
\bm x_e(u)=S_d^{-1}\left[\Delta\overline{\bm d}^{\ddagger}_{e}(u)-\Delta\overline{\bm d}^{\ddagger}_{e,{\rm ref}}\right].
\end{equation}
Within a preregistered descriptor neighborhood, the central predictor and its frozen curvature certificate are
\begin{equation}
\boxed{\beta\widehat{\Delta G}^{\ddagger}_{e}(u)=c_e+\bm a_e^{\mathsf T}\bm x_e(u),\qquad |\rho^{(2)}_e(\bm x_e)|\le \tfrac12 M_e\|\bm x_e\|_2^2.}
\label{eq:linear-main}
\end{equation}
The true molecular surface may be nonlinear. The claim is narrower: $M_e$, the descriptor list, scaling matrix $S_d$, coefficients, uncertainty model, transmission-coefficient rule, curvature tolerance, and applicability threshold are frozen from the restricted training panel before held-out sequence or dwell data are examined. The remainder bound widens the predictive interval but never shifts the held-out mean. Failure to certify the bound, or exceeding either frozen curvature or domain threshold, is an out-of-domain result, not a license for nonlinear repair. A nonlinear model may be reported only as a separately trained comparator and cannot rescue failure of Eq.~\eqref{eq:linear-main}.

Conditional on unresolved coordinates $q_t$, the edge hazard is
\begin{equation}
h_{e}(t\mid q,u)=\kappa_e(q_t,u)\frac{k_BT}{h_{\rm P}}\exp[-\beta\widehat{\Delta G}^{\ddagger}_{e}(q_t,u)].
\label{eq:hazard-main}
\end{equation}
Here $h_{\rm P}$ denotes Planck's constant. The default model uses independently calculated reactive-flux corrections or a preregistered branch rule for $\kappa_e$; the same electronic descriptor is not fitted once in the barrier and again in the prefactor. For two competing exits $b_1,b_2$ from the same register state, evaluated at the same resolved $q_t$, clock time, and perturbation, the common transition-state prefactor scale cancels and Eq.~\eqref{eq:hazard-main} gives the pointwise identity
\begin{equation}
\boxed{\ln\frac{h_{\alpha b_1}}{h_{\alpha b_2}}
=\ln\frac{\kappa_{\alpha b_1}}{\kappa_{\alpha b_2}}
-\beta\!\left(\widehat{\Delta G}^{\ddagger}_{\alpha b_1}-\widehat{\Delta G}^{\ddagger}_{\alpha b_2}\right).}
\label{eq:choice-time-main}
\end{equation}
Thus a barrier contrast that changes nucleotide discrimination simultaneously constrains the exit-rate scale. After hidden-state averaging, integrated choice odds need not equal the pointwise ratio, but choice and timing must still be generated from the same frozen barrier and prefactor family.

The edge kernel
\begin{equation}
\psi_{\alpha\beta}(\tau\mid u)=\left\langle h_{\alpha\beta}(\tau\mid q,u)\exp\!\left[-\int_0^\tau h_\alpha(s\mid q,u)\dd s\right]\right\rangle_{q\mid\alpha,u}
\label{eq:kernel-main}
\end{equation}
therefore links one molecular parameterization to both branch selection and waiting time. Hidden trajectories may generate multiexponential or nonexponential timing, but the exponential Markov kernel remains the null model \cite{English2006,Min2005,Kovalev2009,Maier2026,Frydel2026}.

\section*{A protocol gate before cycle spectroscopy}
For a stationary condition, let $\widehat{\bm\psi}(s)$ be the Laplace-transformed transient event kernel. Where $|z|\rho[\widehat{\bm\psi}(s)]<1$,
\begin{equation}
\boxed{\mathcal Z_{\rm evt}(s,z)=\det[\bm I-z\widehat{\bm\psi}(s)]^{-1}=\prod_{p\in\mathcal P}\left[1-z^{\ell_p}\widehat w_p(s)\right]^{-1}.}
\label{eq:zeta-main}
\end{equation}
The factors represent productive AC writing, A- and C-repeat errors, pause--recovery loops, and other recurrent reaction cycles; derivatives with respect to $s$ generate cycle-time cumulants \cite{ArtinMazur1965,BowenLanford1970,Ruelle1976,CvitanovicEckhardt1991,Bogomolny1992,Baladi1998}. In the minimal two-state limit, a negative subleading eigenvalue may be written $\lambda_{\rm alt}=r_{\rm alt}e^{i\pi}$; this is only a period-2 spectral lemma, not a coherent molecular phase. In stationarity, for any centered binary state observable $y_n$ of this two-state chain,
\begin{equation}
\boxed{C(k)\equiv\langle y_n y_{n+k}\rangle=C(0)\lambda_{\rm alt}^{k},\qquad
\xi_{\rm alt}=-\frac{1}{\ln|\lambda_{\rm alt}|},}
\label{eq:alternating-corr-main}
\end{equation}
for $0<|\lambda_{\rm alt}|<1$. Hence $\lambda_{\rm alt}<0$ predicts directly measurable sign-alternating sequence correlations with a fixed decay length $\xi_{\rm alt}$, without invoking a coherent phase.

A time-dependent experiment must pass a regime gate before Eq.~\eqref{eq:zeta-main} is used. A sufficient coarse-graining condition is $\tau_{\rm fast}\ll\tau_{\rm dwell}$ for variables declared fast, and a sufficient protocol condition is $\tau_{\rm mem}\ll\tau_{\rm protocol}$. More generally, an instantaneous-generator approximation requires
\begin{equation}
\epsilon_{\rm ad}=\operatorname*{ess\,sup}_{t\in I_{\rm reg}}\frac{\|\partial_tL(t)\|}{\gamma(t)^2}\ll1,
\label{eq:adiabatic-main}
\end{equation}
where $I_{\rm reg}$ is a preregistered union of differentiable protocol segments on which the chosen gap estimator remains defined and nonzero \cite{Gu2026}. An unresolved chemical-commitment jump or gap closure is not silently discarded; that window is excluded from the instantaneous approximation and represented by the full nonstationary kernel. If these conditions fail, the correct object is the two-time kernel $\psi_{\alpha\beta}(t_0,\tau)$ and the full age-structured tilted propagator
\begin{equation}
U_\chi(T,0)=\mathcal T\exp\!\left[\int_0^T L_\chi(t)\dd t\right],
\label{eq:prop-main}
\end{equation}
not a stationary Laplace determinant. The kernel is defined after preregistered event projection and temporal resolution; sub-resolution electronic-to-conformational transients are integrated into its normalized waiting-time law, not interpreted separately.

Under any differentiable invertible change of hidden-state representation, the full propagator transforms covariantly. When physical chemical-commitment surfaces, event-injection maps, and boundary-flux operators are transformed consistently, the contracted first-event kernel is representation invariant; an uncontracted boundary kernel is endpoint covariant. For a closed representation, its event-level monodromy spectrum is therefore invariant. These statements are exact and do not require adiabatic driving. Adiabaticity licenses an instantaneous-eigenspace approximation but does not force a geometric current to vanish: slowly driven stochastic pumps can retain a geometric contribution. Such a term is admitted only with a resolved two-parameter closed protocol, a controlled spectral gap, a counting observable, and dynamic-versus-geometric and loop-reversal controls \cite{SinitsynNemenman2007,Rahav2008}.

\section*{Response geometry makes mechanism classes falsifiable}
Let $\bm O$ contain polymerization flux, sequence selectivity, alternating-mode strength, direction-specific dwell statistics, processive survival, and validated thermodynamic estimators, and let $\bm u$ contain independently controlled perturbations. Define
\begin{equation}
\mathcal R=\frac{\partial\bm O}{\partial\bm u},\qquad \widetilde{\mathcal R}=W_O^{1/2}\mathcal R W_u^{1/2}.
\label{eq:response-main}
\end{equation}
The weights are fixed from measurement and perturbation uncertainty. Mechanism classes are preregistered as low-dimensional response subspaces, tested by cross-validated singular-value ratios, normalized $2\times2$ minors, null-space projections, and rescue \cite{Mochizuki2015,BalsaCanto2018,Hosoda2024}. A successful initiation bypass should largely remove the Tyr-sensitive direction from conditional elongation statistics. Recognition-site charge/geometry series should produce complementary sequence and dwell responses. Broad chemistry perturbations should resemble a common barrier shift only if branch effects are substrate independent. Interface perturbations should preferentially alter survival and length tails. Failure of these predictions rejects the proposed decomposition rather than inviting a new fit.

The independent DRT3 construct creates a fourth transfer tier. After sequence and structural alignment are frozen, mapped perturbations should occupy comparable response directions even if absolute rates and residue numbers differ. The test concerns transported mechanism geometry, not post hoc residue-name equivalence. Its concordance must also exceed a preregistered chemistry/geometry-matched null ensemble of admissible alternative mappings; the construction and tail probability are specified in the Supplementary Theory.

\section*{Blind validation, thermodynamic discipline, and scope}
One frozen molecular parameterization must be learned from restricted nucleotide subsets, single-turnover or rapid-quench kinetics, pyrophosphate measurements, and prespecified QM/MM profiles. It must then predict without condition-specific refitting: (i) all-dNTP sequence odds and primitive-cycle weights; (ii) direction-specific dwell distributions and cycle times; (iii) product-length tails, held-out mutations, metal/pH/isotope effects, and initiation rescue; and (iv) cross-construct response directions. Comparators include independent kinetic channels, an unconstrained sequence Markov model, a static semi-Markov register, a dynamic hidden semi-Markov register, and, only when the protocol gate is passed, a driven extension. If a finite continuous-time Markov/phase-type hidden realization of dimension $d$ is claimed, its transfer kernel must additionally satisfy the associated Cayley--Hamilton recurrence and Hankel-rank bound; failure rejects that finite-realization subclass without invalidating the more general semi-Markov formulation.

Primitive-cycle affinities and current statistics may organize chemical cost, while waiting-time asymmetry may provide lower bounds on hidden entropy production \cite{Schnakenberg1976,Seifert2012,AndrieuxGaspard2008,Skinner2021,Ertel2022,Ertel2024,FritzSeifert2026}. These estimators are reported only under their proven reversal, direction--time, and resolution assumptions. No minimum-dissipation optimum, zero-loss operation, or universal entropy--memory extremum is assumed.

The central proposition is deliberately narrow: a finite protein architecture may encode a renewable conditional free-energy rule. Its credibility rests on a locked electronic-to-event map and cross-observable prediction. Unstable barrier coefficients, out-of-domain extrapolation, sequence--time disagreement, failed rank/minor/null-space/rescue constraints, failed construct transport, or equivalent performance by a simpler conventional model would reduce the framework to a descriptive reparameterization.

\begin{figure}[H]
\centering
\includegraphics[width=0.72\textwidth]{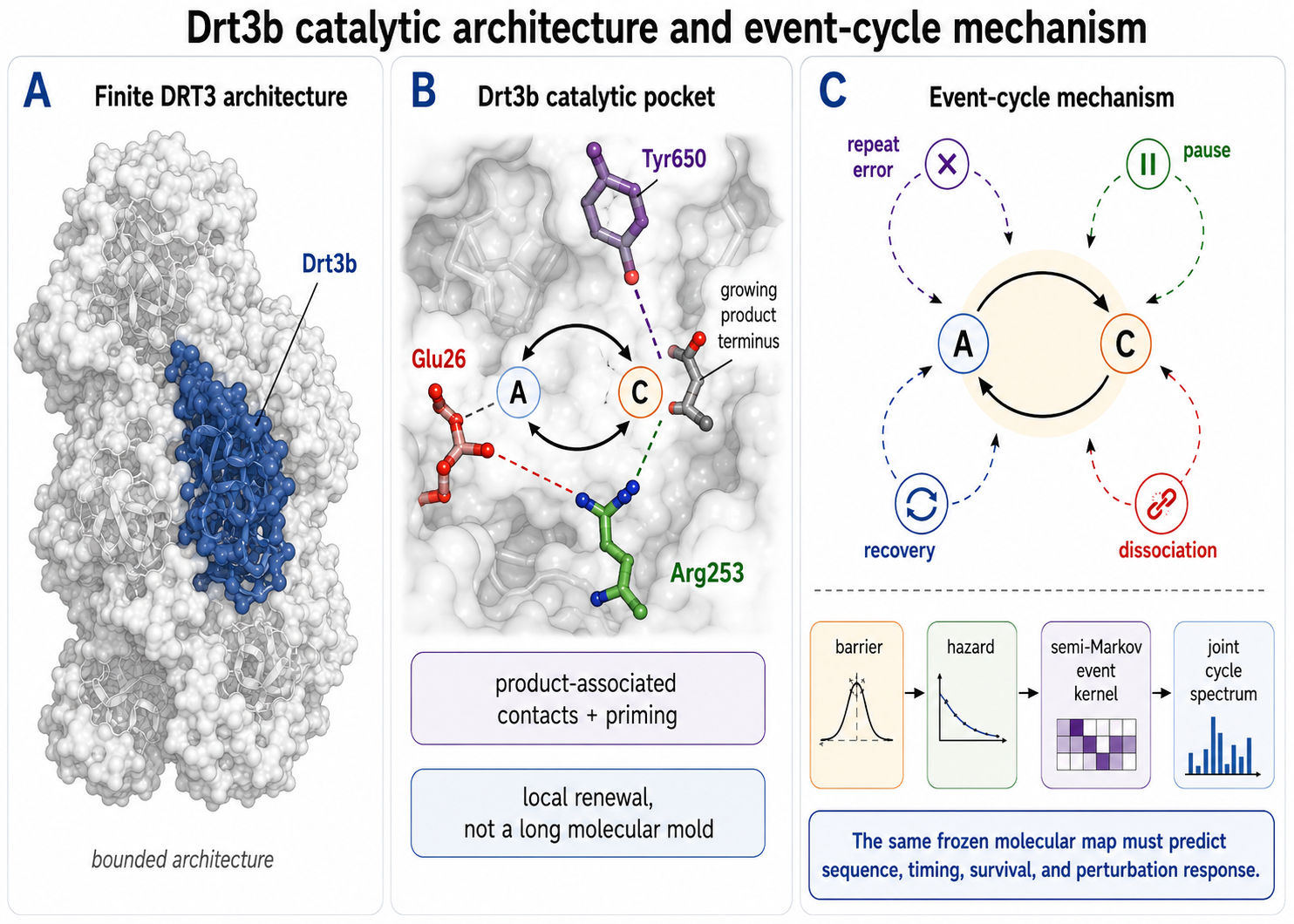}
\caption{Drt3b catalytic architecture and event-cycle mechanism. (A) Finite DRT3 architecture with Drt3b highlighted. (B) A bounded Tyr650/Glu26/Arg253 pocket contacts the growing product terminus. (C) Productive A/C alternation is embedded in repeat-error, pause, recovery, and dissociation cycles. Conceptual schematic; not a coordinate-exact rendering.}
\label{fig:architecture}
\end{figure}

\textbf{Acknowledgments and disclosures.} During preparation, the author used an OpenAI language model for literature organization, conceptual stress-testing, consistency checks, figure-layout development, and language refinement. It was neither an author nor a citable scientific source. All citations were independently verified. The author independently conceived the scientific premises, critically reviewed the manuscript, and accepts full responsibility for its accuracy, integrity, originality, and content.

\textbf{Competing interests:} The author declares no competing interests.

\textbf{Data and materials availability:} No new experimental data were generated. Structural coordinates discussed here are available from the Protein Data Bank under 9Z6Y and 9Z6Z.

\printbibliography[title={References}]
\end{document}


\renewcommand{\thefigure}{S\arabic{figure}}
\renewcommand{\thetable}{S\arabic{table}}
\begin{center}
{\Large\bfseries Supplementary Theory}\par
\vspace{0.18em}
{\large for ``A regenerating free-energy register for protein-templated period-2 DNA synthesis by Drt3b''}\par
\vspace{0.48em}
Wei-Wei Zhang, MD, PhD\par
Adventin Inc., San Diego, California, USA\par
\end{center}

\section*{S1. Empirical anchors, independent convergence, and claim discipline}
The directly supported anchors from Deng et al. are a $6{:}6{:}6$ $D_3$-symmetric DRT3 assembly; RNA-templated poly(GT) synthesis by Drt3a; protein-primed poly(AC) synthesis by Drt3b without a nucleic-acid template; elongating and resting structures 9Z6Y and 9Z6Z at 2.6~\AA; product-associated Glu26--dA and Arg253--dC contacts; and covalent priming from Tyr650 \cite{Deng2026,PDB9Z6Y,PDB9Z6Z}. A reported small $C_\alpha$ superposition supports a compact architecture but does not establish rigidity of all catalytic coordinates or exclude transient local substates.

Wang et al. independently report DRT3b protein-primed, amino-acid-gated poly(dCdA) synthesis in a distinct construct, with E22, R241, and Y666 implicated in gating, specificity, and priming \cite{Wang2026DRT3}. The construct and numbering differ from Deng et al. Therefore:
\begin{enumerate}
\item sequence alignment, structural superposition, and active-site correspondence must be completed without reference to the proposed response model;
\item the alignment map must be frozen before response directions are compared;
\item labels such as ``E26 corresponds to E22'' are not used unless supported by that frozen alignment;
\item cross-construct agreement is tested on descriptor changes and response directions, not on residue numbers alone.
\end{enumerate}

The following remain model objects: a renewable free-energy register, metastable hidden states, electronically conditioned hazards, a semi-Markov kernel, primitive-cycle factors, low-dimensional response geometry, and cross-construct transport. None is validated by a preferred phase, integer, code size, acronym, or external theory. Each object must improve held-out prediction or be removed.

\section*{S2. Conditional quantum mechanics/molecular mechanics (QM/MM) free energy and a bounded first-order barrier map}
For register state $r$, nucleotide $b$, perturbation $u$, reaction coordinate $\xi$, and environmental coordinates $Y$,
\begin{equation}
G_{r,b}(\xi;u)=-k_BT\ln\int \dd Y\,e^{-\beta H_{\rm QM/MM}(\xi,Y;r,b,u)}+C,
\label{eq:pmf}
\end{equation}
with activation free energy
\begin{equation}
\Delta G_{r,b}^{\ddagger}(u)=G_{r,b}(\xi^{\ddagger};u)-G_{r,b}(\xi_R;u).
\label{eq:barrier}
\end{equation}
Collective variables must resolve bond formation and breaking and any solvent, metal, proton, product-end, or conformational coordinate that materially changes the committor. Convergence is audited against QM-region size, electronic-structure level, collective-variable choice, independent trajectory blocks, and restraint or path protocol \cite{Warshel1976,TorrieValleau1977,Pan2023,LiChan2024}.

For a sampled configuration $X$, define a method-declared attribution vector
\begin{equation}
\bm d_{r,b}(X)=
\begin{pmatrix}
E_{\rm elst}&E_{\rm exch}&E_{\rm ind/pol}&E_{\rm disp}&E_{\rm CT}&
\bm E_{\rm loc}\!\cdot\!\Delta\bm\mu^{\ddagger}&q_{\rm transfer}&
{\rm CN}_{\rm Mg}&n_{\rm HB}&\zeta_{\rm water}
\end{pmatrix}^{\!\mathsf T}.
\label{eq:descriptor}
\end{equation}
The raw vector intentionally contains heterogeneous physical components: energetic terms, a field--dipole coupling, charge transfer, and dimensionless coordination/contact observables. Before any linear transfer or projection, the preregistered diagonal scaling operator $S_d$ maps them into one homogeneous dimensionless coordinate system. Here $\zeta_{\rm water}$ is a prespecified hydration-network observable, not an assumed local dielectric constant. Ensemble differences are
\begin{equation}
\Delta\overline{\bm d}_{e}^{\ddagger}(u)=
\langle\bm d_e\rangle_{\rm TS,u}-\langle\bm d_e\rangle_{\rm R,u}.
\label{eq:descriptor-difference}
\end{equation}
Free-energy profile decomposition and QM/MM energy-decomposition analysis may diagnose electrostatic, polarization, van der Waals, exchange, and related contributions, but these are not additive replacements for Eq.~\eqref{eq:barrier} \cite{Pan2023,Xiong2024}.

All scaling is learned from the restricted training panel only:
\begin{equation}
\bm x_e(u)\equiv\Delta\widetilde{\bm d}_{e}^{\ddagger}(u)=
S_d^{-1}\left[\Delta\overline{\bm d}_{e}^{\ddagger}(u)-
\Delta\overline{\bm d}_{e,{\rm ref}}^{\ddagger}\right].
\label{eq:scaled-descriptor}
\end{equation}
Let $g_e(\bm x)\equiv\beta\Delta G_e^{\ddagger}(\bm x)$ denote the conditional barrier restricted to the preregistered descriptor manifold. If $g_e$ is twice differentiable on a declared convex neighborhood $\mathcal N_e$ containing the segment $\{t\bm x:0\le t\le1\}$, then
\begin{equation}
g_e(\bm x)=c_e+\bm a_e^{\mathsf T}\bm x+\rho_e^{(2)}(\bm x),
\label{eq:true-local-barrier}
\end{equation}
where
\begin{equation}
\rho_e^{(2)}(\bm x)=
\int_0^1(1-t)\,\bm x^{\mathsf T}H_e(t\bm x)\bm x\,\dd t .
\label{eq:taylor-remainder}
\end{equation}
The claimed central predictor is only the first-order term:
\begin{equation}
\boxed{\beta\widehat{\Delta G}_{e}^{\ddagger}(u)=
 c_e+\bm a_e^{\mathsf T}\bm x_e(u).}
\label{eq:linear-barrier}
\end{equation}
This does not assert that the physical potential-energy or free-energy surface is globally linear. If the spectral norm of the descriptor-manifold Hessian obeys
\begin{equation}
\sup_{\bm y\in\mathcal N_e}\|H_e(\bm y)\|_2\le M_e,
\end{equation}
then
\begin{equation}
\boxed{|\rho_e^{(2)}(\bm x)|\le
\tfrac12 M_e\|\bm x\|_2^2.}
\label{eq:curvature-bound}
\end{equation}
In practice, a conservative upper confidence bound $\widehat M_e^{\rm up}$ is frozen from a prespecified training-only curvature audit: independent QM/MM points along declared finite-difference stencils or perturbation paths, replicate trajectory blocks, and a stated numerical-error budget. This is an empirical certificate conditional on the audited neighborhood, not a proof of global convergence. Its validity requires the full line segment $\{t\bm x:0\le t\le1\}$ to remain inside the compact closure of the differentiable audited region. A cusp, fold, bifurcation, loss of differentiability, or departure from that closure triggers an out-of-domain result. The bound enlarges predictive uncertainty; it is never fitted as a mean correction on held-out data.

No quadratic interaction tensor, descriptor cross-term, neural network, kernel regressor, or condition-specific correction is admitted into the claimed Tiers I--IV. A nonlinear model may be included only as an explicitly separate comparator trained and frozen under the same data split. Its success does not rescue the first-order generating-register claim.

\subsection*{S2.1 Freeze ledger}
Before held-out sequence or timing data are inspected, the following ledger is timestamped:
\begin{enumerate}
\item state and edge definitions;
\item descriptor names, computational definitions, units, signs, and any permitted coordinate transformation;
\item $S_d$, reference state or reference calculation, and missing-data rules;
\item coefficients $c_e,\bm a_e$ and their covariance;
\item curvature-audit stencil, $\widehat M_e^{\rm up}$, numerical-error allowance, and tolerance $\delta_{G,e}$;
\item the rule for $\kappa_e$ in Eq.~\eqref{eq:hazard};
\item uncertainty propagation and posterior-predictive simulation procedure;
\item descriptor-distance and curvature-budget applicability thresholds;
\item training, validation, and held-out perturbation identities;
\item null comparators and failure thresholds;
\item if a finite realization is claimed, its candidate dimension $d$, recurrence/rank estimator, and numerical tolerance;
\item for cross-construct transport, the admissible null-map ensemble $\Pi_0$ and its sampling rule.
\end{enumerate}

\subsection*{S2.2 Dual domain-of-applicability gate}
Let $\bm\mu_d$ and $\Sigma_d$ be training-panel descriptor moments. A prespecified extrapolation score is
\begin{equation}
D_{\rm app}^2(u)=
\left(\bm x_e(u)-\bm\mu_d\right)^{\mathsf T}
\Sigma_d^{+}
\left(\bm x_e(u)-\bm\mu_d\right),
\label{eq:app-domain}
\end{equation}
where $\Sigma_d^{+}$ is a declared regularized pseudoinverse. Because descriptor distance alone is not a Taylor-convergence theorem, a second gate uses the frozen curvature budget:
\begin{equation}
C_e(u)=
\frac{\tfrac12\widehat M_e^{\rm up}\|\bm x_e(u)\|_2^2}
{\delta_{G,e}}.
\label{eq:curvature-budget}
\end{equation}
A claimed first-order transfer is in domain only if both
\begin{equation}
D_{\rm app}^2(u)\le D_{{\rm thresh},e}^2,
\qquad C_e(u)\le1.
\label{eq:dual-domain}
\end{equation}
All quantities are frozen before the held-out outcome is inspected. Failure of either gate, loss of local differentiability, or approach to a cusp, fold, or bifurcation that invalidates the frozen curvature certificate is classified as out of domain. It is rejected rather than repaired by rescaling, neighborhood expansion, an added nonlinear term, or residual fitting. A separately trained nonlinear comparator may be reported for comparison \cite{PlateroRochart2023}, but it cannot rescue failure of Eqs.~\eqref{eq:app-domain}--\eqref{eq:dual-domain}.

\subsection*{S2.3 No double counting of barrier and prefactor}
Electronic descriptors enter the primary model through Eq.~\eqref{eq:linear-barrier}. The diagonal scaling operator $S_d$ maps heterogeneous raw observables into dimensionless coordinates only for cross-descriptor comparison, attribution, and transfer; it does not privilege a unique valence, bonding, hybridization, or orbital-mechanism interpretation. The selected electronic-structure protocol, scaling parameters, and transmission rule are nevertheless frozen before validation. A descriptor component is not fitted independently into both the barrier and the transmission coefficient using the same dwell data; a prefactor-dependent alternative is a separate comparator. The framework does not infer mechanism from elemental labels alone or from a prespecified symmetry, angle, local permittivity, orbital-hybridization narrative, atomic index, or registry label; only frozen electronic-structure observables and held-out predictions are admissible.

\subsection*{S2.4 Reference-conditioned descriptors and coordinate-invariant prediction}
Atomic or fragment quantities may enter only as declared reference calculations or measurements. They do not supply a universal ``elemental energy'' for a catalytic site. A physically relevant descriptor is instead a matched environmental displacement, for example
\begin{equation}
\delta d_j(X,u)=d_j^{\rm pocket}(X,u)-d_j^{\rm ref}(Z,C,\mu),
\label{eq:environmental-residual}
\end{equation}
where the reference specifies nuclear charge or fragment identity $Z$, electronic configuration or chemical state $C$, computational protocol $\mu$, and any other quantities needed to make the subtraction well defined. Protein, solvent, metal, protonation, and field effects remain in the pocket term and ultimately in the sampled PMF of Eq.~\eqref{eq:pmf}. Reference normalization is therefore a declared baseline for attribution, not an alternative source of activation free energy.

The linear predictor is invariant under any frozen invertible linear change of descriptor coordinates. If
\begin{equation}
\bm x'_e=A\bm x_e,\qquad \bm a'_e=A^{-\mathsf T}\bm a_e,
\end{equation}
then
\begin{equation}
\boxed{(\bm a'_e)^{\mathsf T}\bm x'_e=\bm a_e^{\mathsf T}\bm x_e.}
\label{eq:descriptor-coordinate-invariant}
\end{equation}
Thus individual coefficient magnitudes are representation dependent unless a descriptor basis and normalization are fixed. Mechanistic credit is assigned to invariant predicted barriers, directional derivatives, and held-out responses, not to a preferred coordinate label. The admissible descriptor basis, scaling, and any allowed coordinate transformations are frozen in the ledger before validation.

\section*{S3. Electronically conditioned hazards and stationary or nonstationary kernels}
Let $q_t$ collect unresolved coordinates conditional on state $\alpha$ and perturbation $u$. For edge $e\equiv\alpha\to\beta$,
\begin{equation}
\boxed{h_{\alpha\beta}(a,t_0\mid q,u)=
\kappa_{\alpha\beta}(q_{t_0+a},u)
\frac{k_BT}{h_{\rm P}}
\exp[-\beta\widehat{\Delta G}_{\alpha\beta}^{\ddagger}(q_{t_0+a},u(t_0+a))].}
\label{eq:hazard}
\end{equation}
Here $h_{\rm P}$ denotes Planck's constant. The first time argument $a$ is event age and $t_0$ is the clock time of entry into state $\alpha$. The total exit hazard includes productive, error, pause, off-pathway, and absorbing exits:
\begin{equation}
h_\alpha(a,t_0\mid q,u)=\sum_{\gamma\in\mathcal S}h_{\alpha\gamma}(a,t_0\mid q,u)+h_{\alpha\to{\rm off}}(a,t_0\mid q,u).
\end{equation}
The nonstationary edge kernel is
\begin{equation}
\boxed{\psi_{\alpha\beta}(t_0,\tau\mid u)=
\left\langle h_{\alpha\beta}(\tau,t_0\mid q,u)
\exp\!\left[-\int_0^\tau h_\alpha(s,t_0\mid q,u)\dd s\right]
\right\rangle_{q\mid\alpha,t_0,u}.}
\label{eq:two-time-kernel}
\end{equation}
With all outcomes included,
\begin{equation}
\sum_\beta\int_0^\infty\psi_{\alpha\beta}(t_0,\tau\mid u)\dd\tau=1.
\end{equation}
The event kernel is defined only after preregistered event projection and temporal resolution; sub-resolution inertial, vibrational, electronic, hydration, and conformational transients are integrated into the normalized projected waiting-time law rather than assigned independent event states.

For a stationary condition, clock-time dependence disappears and
\begin{align}
T_{\alpha\beta}(u)&=\int_0^\infty\psi_{\alpha\beta}(\tau\mid u)\dd\tau,\\
f_{\alpha\beta}(\tau\mid u)&=\psi_{\alpha\beta}(\tau\mid u)/T_{\alpha\beta}(u),\qquad T_{\alpha\beta}>0.
\end{align}
Thus the same locked barrier model controls both what is written and when it is written. More sharply, for two competing exits $b_1,b_2$ from the same state, evaluated pointwise at the same $q$, $a$, $t_0$, and $u$,
\begin{equation}
\boxed{\ln\frac{h_{\alpha b_1}}{h_{\alpha b_2}}
=\ln\frac{\kappa_{\alpha b_1}}{\kappa_{\alpha b_2}}
-\beta\left(\widehat{\Delta G}^{\ddagger}_{\alpha b_1}
-\widehat{\Delta G}^{\ddagger}_{\alpha b_2}\right).}
\label{eq:choice-time-lock}
\end{equation}
The identity follows directly from Eq.~\eqref{eq:hazard}; it is not an additional fitted law. Hidden-state averaging and competition through the total survival factor generally prevent integrated branch odds from reducing to this pointwise ratio. Nevertheless, a perturbation that changes the barrier contrast cannot be fitted independently in sequence and timing channels: the same frozen barrier/prefactor family must pass both.

The Markov null is
\begin{equation}
\psi^{\rm M}_{\alpha\beta}(\tau)=T_{\alpha\beta}k_\alpha e^{-k_\alpha\tau}.
\end{equation}
A semi-Markov model is retained only when direction-dependent or nonexponential waiting-time structure improves held-out prediction after missed-event, finite-resolution, change-point, and censoring controls \cite{English2006,Min2005,Kovalev2009,Maier2026}.

\section*{S4. Stationary joint sequence--time cycle determinant}
For a stationary transient kernel, define
\begin{equation}
\widehat\psi_{\alpha\beta}(s)=\int_0^\infty e^{-s\tau}\psi_{\alpha\beta}(\tau)\dd\tau.
\end{equation}
For $|z|\rho[\widehat{\bm\psi}(s)]<1$,
\begin{align}
\mathcal Z_{\rm evt}(s,z)
&=\exp\!\left[\sum_{n\ge1}\frac{z^n}{n}\Tr\widehat{\bm\psi}(s)^n\right]\nonumber\\
&=\det[I-z\widehat{\bm\psi}(s)]^{-1}\nonumber\\
&=\prod_{p\in\mathcal P}\left[1-z^{\ell_p}\widehat w_p(s)\right]^{-1},
\label{eq:event-zeta}
\end{align}
where
\begin{equation}
\widehat w_p(s)=\prod_{(\alpha\to\beta)\in p}\widehat\psi_{\alpha\beta}(s).
\end{equation}
The product is a dynamical Euler product over primitive reaction cycles, not a number-theoretic prime product \cite{ArtinMazur1965,BowenLanford1970,Ruelle1976,CvitanovicEckhardt1991,Bogomolny1992,Baladi1998}.

At $s=0$,
\begin{equation}
\mathcal Z_{\rm evt}(0,z)=\det(I-zT)^{-1}.
\end{equation}
For cycle $p$,
\begin{align}
-\partial_s\ln\widehat w_p(s)|_{s=0}&=\sum_{e\in p}\langle\tau_e\rangle,\\
\partial_s^2\ln\widehat w_p(s)|_{s=0}&=\sum_{e\in p}\Var(\tau_e)
\quad\text{under edge-time independence}.
\end{align}
Mixed cumulants are retained if edge times are dependent. Productive AC cycles, repeat-error one-cycles, pause--recovery loops, and other recurrent paths must therefore give mutually consistent sequence and timing statistics.

For the minimal base chain
\begin{equation}
T_{AC}=\begin{pmatrix}\epsilon_A&1-\epsilon_A\\1-\epsilon_C&\epsilon_C\end{pmatrix},
\qquad
\lambda_0=1,\quad\lambda_{\rm alt}=\epsilon_A+\epsilon_C-1.
\end{equation}
If $\epsilon_A+\epsilon_C<1$,
\begin{equation}
\lambda_{\rm alt}=-r_{\rm alt}=r_{\rm alt}e^{i\pi}.
\end{equation}
This is only a period-2 spectral lemma. It is not a coherent phase, conformational rotation, or topological transition. In stationarity, for any centered binary state observable $y_n$ of this two-state chain, the zero-mean subspace is one dimensional, so
\begin{equation}
\boxed{C(k)\equiv\E[y_ny_{n+k}]=C(0)\lambda_{\rm alt}^{k}.}
\label{eq:alternating-correlation}
\end{equation}
For $0<|\lambda_{\rm alt}|<1$, define the alternating correlation length
\begin{equation}
\boxed{\xi_{\rm alt}=-\frac{1}{\ln|\lambda_{\rm alt}|}.}
\label{eq:alternating-correlation-length}
\end{equation}
Thus $\lambda_{\rm alt}<0$ predicts sign-alternating lag correlations with a fixed exponential envelope. This converts the period-2 spectral statement into a direct sequence-level observable without assigning a coherent molecular phase. In an open process, the determinant is formed on the transient substochastic kernel; the leading survival pole controls product-length tails while recurrent factors describe writing before absorption.

\subsection*{S4.1 Finite-realization recurrence and Hankel-rank falsifier}
The general semi-Markov model does not assume a finite Markov realization. If, however, a candidate hidden mechanism is claimed to admit a $d$-dimensional continuous-time Markov or phase-type realization with subgenerator $K$, input vector $\bm b$, and readout $\bm c$, then a scalar transformed kernel has the rational form
\begin{equation}
\widehat\psi(s)=\bm c^{\mathsf T}(sI-K)^{-1}\bm b
=\sum_{n\ge0}m_n s^{-(n+1)},\qquad
m_n=\bm c^{\mathsf T}K^n\bm b,
\label{eq:finite-realization-expansion}
\end{equation}
for sufficiently large $|s|$. Let
\begin{equation}
p_K(\lambda)=\lambda^d+a_{d-1}\lambda^{d-1}+\cdots+a_0
\end{equation}
be the characteristic polynomial of $K$. Cayley--Hamilton then forces the exact recurrence
\begin{equation}
\boxed{m_{n+d}+a_{d-1}m_{n+d-1}+\cdots+a_0m_n=0,\qquad n\ge0.}
\label{eq:finite-recurrence}
\end{equation}
Equivalently, the infinite Hankel matrix $H_{ij}=m_{i+j}$ obeys
\begin{equation}
\boxed{\operatorname{rank}H\le d.}
\label{eq:hankel-rank}
\end{equation}
These are model-order constraints, not a proof that the molecular dynamics is Markovian. The dimension $d$, fitting window, regularization, and numerical tolerance are frozen from training/validation data. A stable violation on held-out dwell-time data rejects the claimed finite-realization subclass; it does not invalidate a genuinely non-rational semi-Markov kernel. Conversely, increasing $d$ after each held-out failure is counted as model repair rather than prediction.

\section*{S5. Exact age-structured propagation and representation covariance}
Introduce the age $a$ since the last event and let $p_\alpha(a,t)$ be the density in state $\alpha$. For a time-dependent protocol,
\begin{equation}
\partial_t p_\alpha(a,t)+\partial_a p_\alpha(a,t)
=-h_\alpha(a,t;u)p_\alpha(a,t),
\label{eq:age-pde}
\end{equation}
with renewal boundary
\begin{equation}
p_\beta(0,t)=\sum_\alpha\int_0^\infty
h_{\alpha\beta}(a,t;u)p_\alpha(a,t)\dd a.
\label{eq:renewal-boundary}
\end{equation}
This enlarged description restores local Markov evolution in state--age space \cite{Frydel2026,Becker2000}.

Write the evolution as $\dot p=L(t)p$. Under a differentiable invertible representation $p(t)=D(t)p'(t)$,
\begin{equation}
\boxed{L'(t)=D^{-1}(t)L(t)D(t)-D^{-1}(t)\dot D(t).}
\label{eq:cov-generator}
\end{equation}
The propagator transforms exactly as
\begin{equation}
U'(t_2,t_1)=D^{-1}(t_2)U(t_2,t_1)D(t_1).
\end{equation}
For a closed representation, $D(T)=D(0)$,
\begin{equation}
\boxed{\det[I-zU'(T,0)]=\det[I-zU(T,0)].}
\label{eq:monodromy-invariant}
\end{equation}
This covariance is an algebraic identity for the full propagator. It does not require slow driving and is not evidence for a physical Berry phase. What requires a time-scale or spectral-gap condition is the replacement of the exact nonstationary propagator by a sequence of instantaneous stationary kernels or eigenspaces.

\subsection*{S5.1 Proposition: covariance of event projection}
Let $U_0(t_2,t_1)$ be the time-ordered \emph{no-event} propagator generated on the live state--age space by the killed generator, before renewal reinjection. Let $\mathcal R_\alpha(t_0)$ inject an event state $\alpha$ into the physical age-zero entrance distribution, and let $\mathcal J_\beta(t)$ be the boundary-flux functional for the first chemical-commitment event of type $\beta$. The operator form of the first-event kernel is
\begin{equation}
\boxed{\Psi_{\beta\alpha}(t_0,\tau)=
\mathcal J_\beta(t_0+\tau)\,
U_0(t_0+\tau,t_0)\,
\mathcal R_\alpha(t_0).}
\label{eq:event-projection}
\end{equation}
Equation~\eqref{eq:event-projection} is the boundary-flux representation of Eq.~\eqref{eq:two-time-kernel}. Under $p(t)=D(t)p'(t)$, transform the three ingredients consistently:
\begin{align}
U_0'(t_2,t_1)&=D^{-1}(t_2)U_0(t_2,t_1)D(t_1),\\
\mathcal R_\alpha'(t_0)&=D^{-1}(t_0)\mathcal R_\alpha(t_0),\\
\mathcal J_\beta'(t)&=\mathcal J_\beta(t)D(t).
\end{align}
The contracted physical kernel is then exactly invariant,
\begin{equation}
\boxed{\Psi_{\beta\alpha}'(t_0,\tau)=
\Psi_{\beta\alpha}(t_0,\tau).}
\label{eq:event-kernel-invariant}
\end{equation}
If internal boundary coordinates are retained rather than contracted, the corresponding boundary operator is endpoint covariant,
\begin{equation}
\bm\Psi'(t_0,\tau)=
D_\partial^{-1}(t_0+\tau)\,
\bm\Psi(t_0,\tau)\,
D_\partial(t_0),
\label{eq:event-endpoint-covariance}
\end{equation}
where $D_\partial$ is the induced representation on the event boundary. For a closed protocol/representation, $D_\partial(T)=D_\partial(0)$, the event-to-event monodromy operator changes by similarity and its spectrum is invariant.

The proposition is conditional on five operational requirements:
\begin{enumerate}
\item stopping surfaces represent the same physical chemical-commitment events before and after reparameterization;
\item $D(t)$ and its induced boundary map are differentiable, bounded, and invertible on the live and event spaces;
\item no-event evolution, entrance injection, and boundary flux are transformed as one system rather than piecemeal;
\item the nonstationary kernel is built from the full time-ordered propagator, not by stitching stationary kernels across a fast protocol;
\item recrossing, missed events, finite temporal resolution, and censoring are included in the event-definition or observation model.
\end{enumerate}
The commitment surfaces, recrossing rule, event-injection/reinjection maps, and any post-event reset rule are frozen before held-out sequence or dwell-time data are inspected. If coarse-graining changes the stopping event itself, or merges paths with different commitment surfaces without an explicit observation model, Eq.~\eqref{eq:event-kernel-invariant} is not claimed. The result removes a representation artifact at event truncation; it does not suppress a physical protocol-dependent geometric current.

\section*{S6. Protocol-regime gate: stationary reduction, adiabatic approximation, or full propagation}
The reviewer-suggested ordering
\begin{equation}
\tau_{\rm relax}^{\rm hydration}\ll\tau_{\rm dwell}\ll\tau_{\rm protocol}
\label{eq:sufficient-timescale}
\end{equation}
is a useful sufficient condition only if hydration has explicitly been assigned to the fast equilibrating sector. It is not universal: hydration, protonation, or metal-coordinate variables may instead belong to the resolved hidden trajectory $q_t$. The manuscript therefore uses two separate gates.

\textbf{Coarse-graining gate.} For variables declared fast,
\begin{equation}
\epsilon_{\rm cg}=\max_\alpha\frac{\tau_{\rm fast,\alpha}}{\tau_{\rm dwell,\alpha}}\ll1.
\label{eq:cg-gate}
\end{equation}
If this fails, the variable is promoted into the hidden state or trajectory rather than averaged as equilibrated noise.

\textbf{Protocol gate.} A simple memory criterion is
\begin{equation}
\epsilon_{\rm prot}=\max_\alpha\frac{\tau_{\rm mem,\alpha}}{\tau_{\rm protocol}}\ll1.
\label{eq:protocol-gate}
\end{equation}
For an instantaneous-generator or eigenspace approximation with relaxation gap $\gamma(t)$, a stronger operator condition is
\begin{equation}
\epsilon_{\rm ad}=\operatorname*{ess\,sup}_{t\in I_{\rm reg}}\frac{\|\partial_tL(t)\|}{\gamma(t)^2}\ll1,
\label{eq:adiabatic-gate}
\end{equation}
where $I_{\rm reg}$ is a preregistered union of protocol segments on which $L(t)$ is differentiable and the chosen gap remains defined; the norm, state space, segment boundaries, and gap estimator are declared in advance \cite{Gu2026}. A short chemical-commitment window is not discarded merely because it is brief. If it contains an unresolved jump or gap closure, that window is excluded from the instantaneous-kernel reduction and represented by the full nonstationary kernel. When Eqs.~\eqref{eq:cg-gate}--\eqref{eq:adiabatic-gate} are satisfied on the regular segments, slowly varying instantaneous kernels and controlled adiabatic corrections may be meaningful. Otherwise, the analysis uses the two-time kernel in Eq.~\eqref{eq:two-time-kernel} and the exact time-ordered propagator
\begin{equation}
U_\chi(T,0)=\mathcal T\exp\!\left[\int_0^T L_\chi(t)\dd t\right].
\label{eq:tilted-propagator}
\end{equation}
The stationary determinant in Eq.~\eqref{eq:event-zeta} is not invoked outside its regime. Equations~\eqref{eq:event-kernel-invariant} and \eqref{eq:event-endpoint-covariance} remain exact when the adiabatic approximation fails because they are statements about the full propagator and consistently transformed event operators.

Adiabaticity does not imply zero geometric pumping. In slowly driven stochastic networks, an integrated current can retain a geometric contribution; its absence requires additional model-specific no-pumping conditions on the driven parameters and rate structure \cite{SinitsynNemenman2007,Rahav2008}. Thus $\epsilon_{\rm ad}\ll1$ licenses an instantaneous spectral expansion but neither creates event-projection covariance nor forces $J_{\rm geom}=0$.

\subsection*{S6.1 Conditional stochastic geometric pumping}
A geometric contribution is considered only when:
\begin{enumerate}
\item at least two independently controlled parameters $\bm\lambda(t)$ trace a resolved closed loop;
\item a counted current and a tilted generator $L_\chi(\bm\lambda)$ are defined;
\item the relevant eigenvalue remains separated by a controlled gap along the loop;
\item dynamic and geometric parts are separated by frequency, orientation, or loop-reversal controls;
\item the result survives a full nonadiabatic propagator calculation.
\end{enumerate}
For dominant left and right eigenvectors, an adiabatic connection may be written
\begin{equation}
\mathcal A_\mu^\chi(\bm\lambda)=
\langle l_0^\chi(\bm\lambda)|\partial_{\lambda^\mu}r_0^\chi(\bm\lambda)\rangle,
\end{equation}
but the connection is not inserted into ordinary probabilities as an arbitrary complex phase \cite{SinitsynNemenman2007,Rahav2008}. No $\pi/2$ step, $\pi$ Berry phase, coherent interference, or topological transition is assumed.

\section*{S7. Weighted sparse response geometry, rank tests, and cross-construct transport}
Let
\begin{equation}
\bm O=(\ln J_{\rm pol},\lambda_{\rm alt},\ln L_{\rm proc},I_{\rm sel},
\langle\tau_{A\to C}\rangle,\langle\tau_{C\to A}\rangle,
F_{AC},\Sigma_{AC},\ldots)^{\mathsf T}
\end{equation}
and let $\bm u$ be independently controlled perturbations. Define
\begin{equation}
\mathcal R=\frac{\partial\bm O}{\partial\bm u},
\qquad
\boxed{\widetilde{\mathcal R}=W_O^{1/2}\mathcal R W_u^{1/2}.}
\label{eq:weighted-response}
\end{equation}
The weights are fixed from measurement and perturbation uncertainty, not chosen to improve rank after inspection. The chain rule may be displayed as
\begin{equation}
\mathcal R=J_{O\leftarrow k}J_{k\leftarrow G}J_{G\leftarrow d}J_{d\leftarrow u},
\label{eq:response-factorization}
\end{equation}
which expresses perturbation $\to$ descriptor $\to$ barrier $\to$ hazard/kernel $\to$ observable \cite{Mochizuki2015,BalsaCanto2018,Hosoda2024}.

For a preregistered mechanism block, let singular values satisfy $\sigma_1\ge\sigma_2\ge\cdots$. A rank-$r$ claim is tested by a frozen ratio such as
\begin{equation}
\rho_r=\frac{\sigma_{r+1}}{\sigma_1},
\end{equation}
with confidence intervals obtained by a declared bootstrap or posterior-predictive scheme. It is not enough to report a visually dominant first component.

For observables $i,j$ and perturbations $k,l$,
\begin{equation}
\mathcal I_{ij}^{kl}=
\mathcal R_{ik}\mathcal R_{jl}-\mathcal R_{il}\mathcal R_{jk}
\end{equation}
may be normalized as
\begin{equation}
\widetilde{\mathcal I}_{ij}^{kl}=
\frac{\mathcal I_{ij}^{kl}}
{\sqrt{(\mathcal R_{ik}^2+\mathcal R_{il}^2)(\mathcal R_{jk}^2+\mathcal R_{jl}^2)}+\varepsilon_0},
\label{eq:normalized-minor}
\end{equation}
where $\varepsilon_0$ is a frozen numerical stabilizer. Only minors associated with a stated one-axis mechanism are predicted to be small.

\small
\setlength{\tabcolsep}{2.8pt}
\begin{longtable}{p{0.18\textwidth}p{0.19\textwidth}p{0.18\textwidth}p{0.19\textwidth}p{0.20\textwidth}}
\caption{Perturbation classes and preregistered discriminators. Rows separate initiation, A- and C-associated recognition, chemistry, processivity, and cross-construct transport. Every direction, threshold, and rescue criterion is frozen before held-out outcomes are inspected.}\label{tab:perturbation-discriminators}\\
\toprule
Perturbation class & Primary target & Sequence/cycle prediction & Timing/processivity prediction & Hard discriminator\\
\midrule
\endfirsthead
\toprule
Perturbation class & Primary target & Sequence/cycle prediction & Timing/processivity prediction & Hard discriminator\\
\midrule
\endhead
Primer Tyr loss or bypass & initiation edge & conditional elongation cycle spectrum preserved after successful bypass if initiation-only & fewer starts; rescued molecules retain elongation dwell laws & persistent Tyr direction after bypass implies renewal or processivity role\\
Glu26 charge/geometry series & A-associated recognition or product registration & A-associated odds and cycle factors shift & direction-specific dwell response & charge/steric series separates recognition from generic damage\\
Arg253 charge/geometry series & C-associated recognition or product registration & complementary C-associated changes & complementary dwell response & tests two-site asymmetry\\
Catalytic residues, metal, pH, D$_2$O & chemistry barrier or prefactor & normalized choice preserved only for branch-uniform shifts & broad time rescaling; isotope or recrossing effect & separates chemistry from discrimination\\
Oligomer/interface & dissociation or off-pathway hazards & local short-cycle spectrum may persist & strong survival and product-length effect & separates local register from processivity\\
Independent construct & aligned homologous mechanism class & mapped descriptor and response direction should transport & absolute rates may differ & alignment frozen before comparison\\
\bottomrule
\end{longtable}
\normalsize
\setlength{\tabcolsep}{6pt}

A null-space rescue is stronger than generic activity recovery. After successful initiation bypass, the Tyr perturbation vector should project weakly onto the conditional elongation-response subspace if Tyr is initiation-only.

\subsection*{S7.1 Cross-construct response transport}
Let $A_{D\to W}$ be the frozen alignment/transport operator from the Deng construct to the Wang construct. For a mechanism-class response direction $\bm r_D$ and an independently measured direction $\bm r_W$, define
\begin{equation}
\cos\theta_{DW}=
\frac{(A_{D\to W}\bm r_D)^{\mathsf T}W\bm r_W}
{\|A_{D\to W}\bm r_D\|_W\,\|\bm r_W\|_W}.
\label{eq:construct-angle}
\end{equation}
The threshold for concordance is preregistered with uncertainty propagation. Residue numbering cannot be altered after the response data are seen. Failure of transported directionality is a direct falsification of cross-construct mechanism generality, not an invitation to realign post hoc.

A positive transport claim must also beat a preregistered matched null. Let $\Pi_0$ be a finite or sampled ensemble of alternative maps that preserve only the declared coarse constraints used to build the null---for example sequence-position class, residue chemistry, local structural neighborhood, or perturbation-class balance---but not the proposed mechanistic correspondence. For each $\pi\in\Pi_0$, compute the same weighted angle $\cos\theta_\pi$. A one-sided tail probability is
\begin{equation}
\boxed{p_{\rm tr}=\Pr_{\pi\sim\Pi_0}
\left[\cos\theta_\pi\ge\cos\theta_{DW}\right].}
\label{eq:transport-null}
\end{equation}
The construction of $\Pi_0$, treatment of map multiplicity, sign conventions, and uncertainty propagation are frozen before the Wang response data are inspected. High $\cos\theta_{DW}$ without separation from this matched null does not count as cross-construct mechanistic transport.

\section*{S8. Cycle thermodynamics and waiting-time entropy bounds}
For an observed trajectory $\omega$ and a precisely defined reversed conjugate trajectory $\widetilde\omega$,
\begin{equation}
\Sigma[\omega]=k_B\ln\frac{\mathbb P[\omega]}{\mathbb P^{\dagger}[\widetilde\omega]},
\end{equation}
so under the correct reverse protocol
\begin{equation}
\left\langle e^{-\Sigma/k_B}\right\rangle=1,
\qquad\langle\Sigma\rangle\ge0
\end{equation}
\cite{LebowitzSpohn1999,Seifert2012}. For a reversible Markov primitive cycle $p$,
\begin{equation}
\mathcal A_p=k_BT\ln\frac{\prod_{e\in p}k_e^+}{\prod_{e\in\bar p}k_e^-},
\qquad
\dot S_{\rm tot}=T^{-1}\sum_pJ_p\mathcal A_p\ge0
\end{equation}
under a valid cycle decomposition \cite{Schnakenberg1976}.

Semi-Markov thermodynamics requires care. Direction--time independence is a substantive condition, not a default; waiting-time irreversibility and hidden paths can add entropy that a Markov rate product misses \cite{AndrieuxGaspard2008,Ertel2022,Maier2026}. Before using a cycle-affinity estimator, the reverse protocol, event resolution, direction--time condition, and censoring treatment are stated.

Waiting-time statistics may supply lower bounds on hidden entropy production, including when full microscopic states are inaccessible \cite{Skinner2021,Ertel2024,FritzSeifert2026}. Define productive-cycle flux
\begin{equation}
J_{AC}=\lim_{t\to\infty}\frac{\E N_{AC}(t)}{t},
\end{equation}
entropy cost per productive cycle
\begin{equation}
\Sigma_{AC}=\frac{\dot S_{\rm tot}}{k_BJ_{AC}},
\end{equation}
and conditional information written per event
\begin{equation}
I_{\rm sel}=\sum_{r,b}P(r,b)\ln\frac{P(b\mid r)}{P_0(b)}.
\end{equation}
An information-cost ratio $\mathcal C_{\rm info}=\Sigma_{AC}/I_{\rm sel}$ may be reported only when both numerator and denominator are identified under the stated coarse-graining. No universal minimum of $\Sigma_{AC}L_{\rm mem}^2$, zero-loss operation, or unique thermodynamic optimum is assumed. Conditions are compared on a Pareto surface of speed, accuracy, processivity, timing, and defensible cost.

\section*{S9. Four-tier no-refit validation, uncertainty, and model comparison}
Training data may include restricted nucleotide subsets, single-turnover rates, rapid-quench and PP$_i$ kinetics, prespecified QM/MM profiles, and a declared subset of perturbations. Held-out data include all-four-dNTP sequences, dwell times, product-length tails, perturbations not used for fitting, and the independent construct.

For parameters $\widehat\vartheta(D_{\rm train})$, evaluate
\begin{equation}
\log p(D_{\rm test}\mid\widehat\vartheta(D_{\rm train}),M)
\end{equation}
without condition-specific refitting. Report predictive density, calibration, interval coverage, effective complexity, sensitivity to state coarse-graining, and all preregistered failures. Compare:
\begin{enumerate}
\item independent Michaelis--Menten or multi-step kinetic channels;
\item an unconstrained sequence Markov chain;
\item a static physical semi-Markov register;
\item a dynamic hidden semi-Markov register;
\item when claimed, a finite Markov/phase-type hidden realization with preregistered dimension $d$ and recurrence/rank tests;
\item the locked linear electronic-to-event model;
\item a nonlinear barrier model only as a distinct comparator;
\item only after the protocol gate, a driven or geometric-pump extension.
\end{enumerate}

The transfer tiers are:
\begin{itemize}
\item \textbf{Tier I}: restricted barriers and rates $\to$ all-dNTP sequence odds and primitive-cycle weights;
\item \textbf{Tier II}: the same frozen parameters $\to$ direction-specific dwell distributions and cycle-time cumulants; when a finite realization is claimed, the same tier also tests Eqs.~\eqref{eq:finite-recurrence} and \eqref{eq:hankel-rank};
\item \textbf{Tier III}: the same frozen parameters $\to$ length tails, weighted response rank/minors, null spaces, held-out mutants, chemistry perturbations, entropy bounds, and rescue;
\item \textbf{Tier IV}: the frozen alignment and transported mechanism map $\to$ independent-construct response direction, together with separation from the preregistered null ensemble in Eq.~\eqref{eq:transport-null}.
\end{itemize}
A model that succeeds only after fitting the held-out observable fails at that tier. Out-of-domain predictions are reported as such and do not count as successful transfer.

\begin{figure}[ht]
\centering
\includegraphics[width=0.99\textwidth]{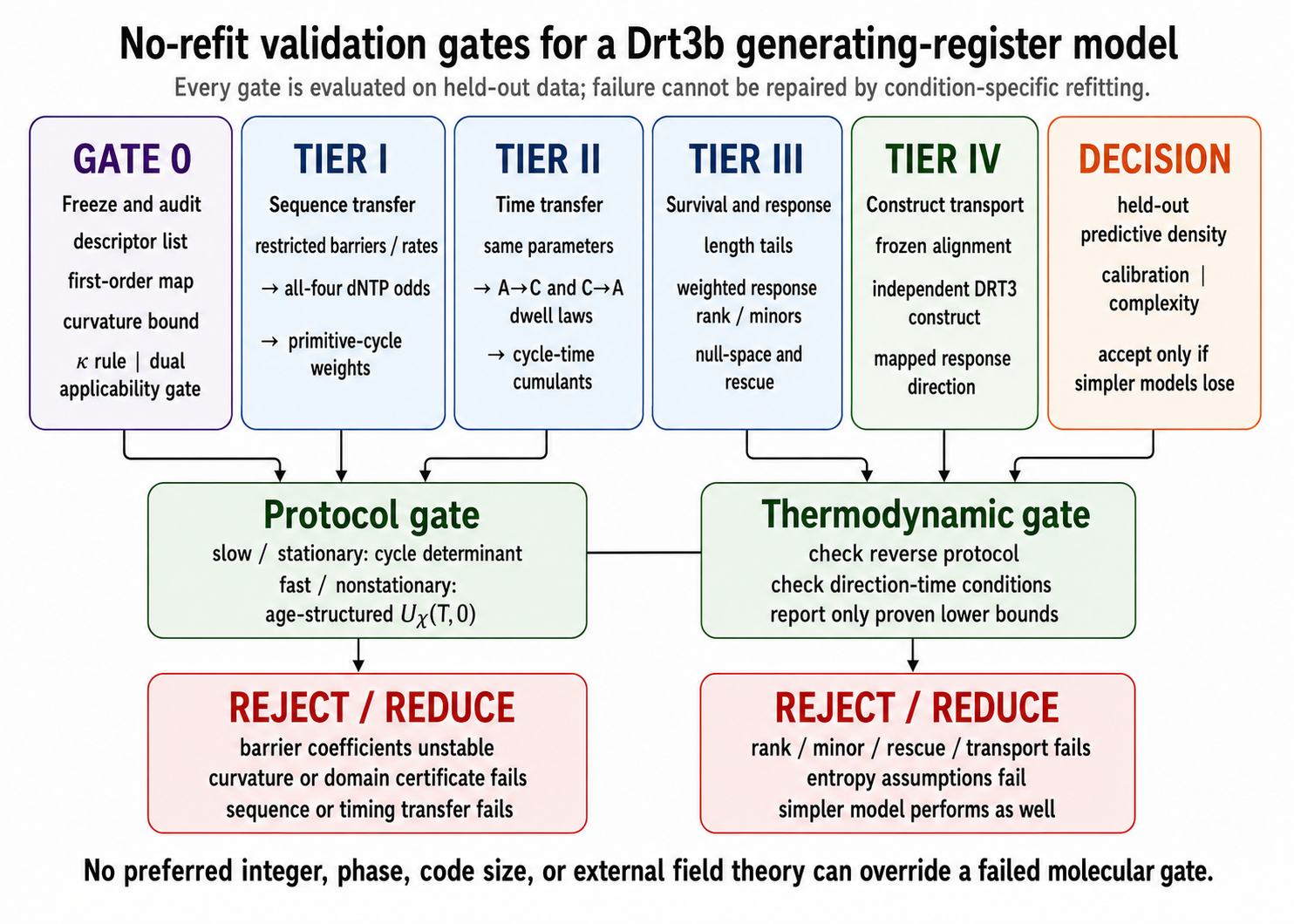}
\caption{No-refit validation gates. Gate 0 freezes the descriptor system, within-domain first-order map, curvature-remainder certificate, transmission rule, uncertainty, and dual applicability gate. Tiers I--IV test sequence, timing, survival/response geometry, and cross-construct transport. Protocol and thermodynamic gates determine which mathematical and energetic estimators are admissible. No preferred integer, phase, code size, or external theory can override a failed molecular gate.}
\label{fig:validation-gates}
\end{figure}

\section*{S10. Experimental and computational program}
\textbf{Phase I: ensemble biochemical closure.} Measure all-four-dNTP and restricted-subset single-turnover kinetics, initiation fraction, PP$_i$ release, sequence reads, product-length distributions, pH, D$_2$O, metal substitutions, and preregistered charge/geometry series.

\textbf{Phase II: electronic and event-time closure.} Compute converged QM/MM PMFs for a limited training panel; perform trajectory-resolved attribution; estimate or calculate transmission corrections independently; and develop time-resolved or single-molecule assays that pair nucleotide identity with dwell time and, where feasible, a local structural observable. If a finite Markov/phase-type hidden realization is proposed, freeze its candidate dimension and test recurrence/Hankel-rank stability on held-out dwell data.

\textbf{Phase III: response-geometry closure.} Estimate $\widetilde{\mathcal R}$; test frozen ranks, singular-value ratios, normalized minors, null spaces, and initiation rescue. Use replicate perturbation magnitudes to verify local linear response rather than infer derivatives from one large perturbation.

\textbf{Phase IV: protocol closure.} Measure protocol and memory time scales. Use the stationary determinant only after Eqs.~\eqref{eq:cg-gate}--\eqref{eq:adiabatic-gate} are supported; otherwise fit the two-time kernel and age-structured propagator. Geometric pumping is tested only with a resolved two-parameter loop and loop-reversal controls.

\textbf{Phase V: independent-construct transport.} Freeze sequence/structure alignment, calculate the aligned descriptor changes, and test response-direction transport without renumbering or model repair. In parallel, freeze a chemistry/geometry-matched null ensemble $\Pi_0$ and require the observed transport angle to separate from that null under Eq.~\eqref{eq:transport-null}.

\section*{S11. Hard falsification and reduction rules}
The generating-register model is rejected or sharply reduced if any of the following occurs:
\begin{itemize}
\item Eq.~\eqref{eq:linear-barrier} is unstable across QM regions, trajectory blocks, or training splits;
\item held-out perturbations lie outside the applicability domain and the claim nevertheless treats them as validated predictions;
\item descriptor coefficients require a quadratic or condition-specific correction to pass the claimed tiers;
\item the same molecular map cannot predict both sequence odds and waiting-time laws, including the pointwise choice--time constraint in Eq.~\eqref{eq:choice-time-lock} within its stated conditioning;
\item a claimed finite Markov/phase-type realization violates the frozen recurrence or Hankel-rank bound on held-out data, or requires increasing its dimension after failure;
\item no stable Markov or semi-Markov event representation exists after resolution and missed-event controls;
\item stationary cycle analysis is applied despite failure of the protocol gate;
\item cycle weights inferred from sequences disagree with rates or dwell data;
\item preregistered rank, minor, null-space, or rescue predictions fail;
\item the frozen cross-construct alignment does not transport mechanism response direction, or its concordance does not exceed the preregistered matched null ensemble;
\item entropy estimators are dominated by hidden-state, reversal, direction--time, censoring, or finite-resolution bias;
\item a simpler conventional model has equal or better held-out predictive density and calibration.
\end{itemize}

\textbf{Scope of inference.} Failure at an earlier tier removes mechanistic credit from downstream tiers; later observables may be reported descriptively but cannot rescue or reparameterize the frozen model. This submission is confined to the molecular observables, perturbations, and prespecified kinetic classes above. No external field equation, cosmological analogy, universal entropy--memory principle, or cross-scale relation enters derivation, validation, or interpretation. The claims stand or fail entirely by the no-refit predictions in Tiers I--IV; no inference beyond that scale is licensed.

\printbibliography[title={References}]